\documentclass[12pt]{emulateapj}
\usepackage{natbib}

\newcommand{\etal}{et~al.}

\newcommand{\hst}{{\it HST}}

\newcommand{\msun}{$M_\odot$}

\def\lri{{LRI}}

\def\spose#1{\hbox to 0pt{#1\hss}}
\def\simlt{\mathrel{\spose{\lower 3pt\hbox{$\mathchar"218$}}
     \raise 2.0pt\hbox{$\mathchar"13C$}}}
\def\simgt{\mathrel{\spose{\lower 3pt\hbox{$\mathchar"218$}}
     \raise 2.0pt\hbox{$\mathchar"13E$}}}

\slugcomment{ApJ, in press} 

\shorttitle{IDCS J1426+3508 at $z = 1.75$}
\shortauthors{Stanford et al.}

\newcommand{\Davis}{1}
\newcommand{\LLNL}{2}
\newcommand{\Missouri}{3}
\newcommand{\CfA}{4}
\newcommand{\UFlorida}{5}
\newcommand{\JPL}{6}
\newcommand{\NOAO}{7}

\begin{document}


\title{IDCS J1426+3508: Discovery of a Massive, IR-Selected Galaxy Cluster at $z = 1.75$}


\author{S.~A.~Stanford,\altaffilmark{\Davis,\LLNL} 
M.~Brodwin,\altaffilmark{\Missouri,\CfA}
Anthony~H.~Gonzalez,\altaffilmark{\UFlorida}
Greg~Zeimann,\altaffilmark{\Davis}
Daniel~Stern,\altaffilmark{\JPL}
Arjun~Dey,\altaffilmark{\NOAO} 
P.~R.~Eisenhardt,\altaffilmark{\JPL}
Gregory~F.~Snyder,\altaffilmark{\CfA} 
and  C.~Mancone\altaffilmark{\UFlorida}
}


\altaffiltext{\Davis}{Department of Physics, University of California, One Shields Avenue, Davis, CA 95616}
\altaffiltext{\LLNL}{Institute of Geophysics and Planetary Physics, Lawrence Livermore National Laboratory, Livermore, CA 94550}
\altaffiltext{\Missouri}{Department of Physics and Astronomy, 5110
  Rockhill Road University of Missouri, Kansas City, MO, 64110}
\altaffiltext{\CfA}{Harvard-Smithsonian Center for Astrophysics, 60 Garden Street, Cambridge, MA 02138}
\altaffiltext{\UFlorida}{Department of Astronomy, University of Florida, Gainesville, FL 32611}
\altaffiltext{\JPL}{Jet Propulsion Laboratory, California Institute of Technology, Pasadena, CA 91109}
\altaffiltext{\NOAO}{NOAO, 950 North Cherry Avenue, Tucson, AZ 85719}



\begin{abstract}

  We report the discovery of an IR-selected massive galaxy cluster
  in the IRAC Distant Cluster Survey (IDCS).  We present
  new data from the Hubble Space Telescope and the W.~M.~Keck
  Observatory that spectroscopically confirm IDCS~J1426+3508 at
  $z=1.75$.  Moreover, the cluster is detected in archival Chandra data as an extended X-ray source, comprising 54 counts after the removal of point sources.  We
  calculate an X-ray luminosity of $L_{0.5-2 keV} = (5.5 \pm 1.2)  \times 10^{44}$ ergs
  s$^{-1}$ within $r = 60$ arcsec ($\sim 1$ Mpc diameter), which implies M$_{200,L_x} = (5.6 \pm 1.6) \times 10^{14}$~M$_\odot$.  
  IDCS~J1426+3508 appears to be an exceptionally massive cluster for its redshift.

\end{abstract}

\keywords{galaxies: clusters: individual --- galaxies: distances and redshifts --- galaxies: evolution}

\section{Introduction}

Galaxy clusters present the opportunity for addressing two main issues in astrophysics.  Cosmological parameters may be constrained with knowledge of the  abundance of clusters if the selection function, redshifts, and masses are adequately known \citep{haiman01, holder01}.   Even individual clusters may be useful if they are sufficiently massive and at sufficiently high redshift because they trace the extreme tail of the cosmological density field \citep{mat00, mort11}.  

Impressive progress has been made over the past decade in finding and characterizing galaxy clusters at $1 < z < 1.5$.   At $z \gtrsim 1.5$  the number of massive, high redshift
clusters that have been identified and confirmed is still very
limited.  The accounting depends on the definitions both of what constitutes a massive cluster and what constitutes confirmation.   Here we assume a high-redshift cluster massive enough to be useful for constraining cosmological parameters has an M$_{200}$ mass (the mass within the region where the cluster overdensity is 200 times the critical density) of at least $1 \times 10^{14}$ M$_\sun$.  Such
objects are the progenitors of present-day clusters with masses of
$\sim 5 \times 10^{14}$~M$_\odot$.  We suggest that confirmation of a cluster candidate requires at least $5$ spectroscopic redshift members within 2 Mpc.

At $z > 1.5$ several systems have been published in the literature.  The estimated masses of the cluster at $z=1.62$ identified by both \citet{pap10} and \cite{tan10}, the cluster at $z=1.75$ reported by \citet{henry10}, and the cluster at $z=2.07$ published by \citet{gobat11} appear to be less than $1 \times 10^{14}$ M$_\odot$.     \citet{santos11} and \citet{fass11} have published spectroscopically confirmed X-ray selected clusters at $z = 1.58$ and $z=1.56$, respectively, both of which appear to have cluster masses of a few$\times 10^{14}$ M$_\odot$, although in the former case only 3 member galaxies have spectroscopic redshifts.  Thus there are only 1--2 galaxy clusters at $z > 1.5$ that are massive, confirmed, and published.   A number of proto--clusters have been identified and confirmed at $z > 2$ \citep{pent00, ven07, capak11}.  While very interesting, these systems appear to be in the very early stages of cluster formation--their masses remain difficult to estimate and their natures difficult to interpret.

The other main astrophysical use for galaxy clusters is to help us understand galaxy evolution.  In particular, they contain the majority of the massive early-type galaxies in the universe, so these environments offer excellent places in which to study such galaxies.  To trace the evolution of massive early-type galaxies over their full lifetime, we must identify and study the {\it precursor} cluster population over a large redshift range.  This kind of archaeology requires the evolutionary precursors to be identified in large, statistically useful samples, which are sensitive down to the group scale at relatively high redshift, since the massive clusters at $z < 1$ are built from groups and low-mass clusters at $z > 1$.  

Identifying and then characterizing cluster samples adequate to these two tasks has been challenging.  Optical methods of finding clusters tend to succeed up to redshifts only slightly beyond unity.  Neither X-ray nor Sunyaev-Zel'dovich (SZ) cluster surveys currently have the sensitivity to reach cluster masses down to $1 \times 10^{14} M_\odot$ at $z \sim 1.5$ and above over appreciable areas.  

The IRAC Shallow Cluster Survey \citep[ISCS,][]{eisenhardt08} originally was designed to create a stellar mass-selected selected sample of galaxy clusters spanning $0 < z < 2$.  The ISCS is drawn from the Spitzer/IRAC Shallow Survey \citep{iss},
which imaged most of the Bo\"otes field in the NOAO Deep Wide-Field Survey \citep{ndwfs99}.   Clusters were identified by searching for 3-dimensional spatial overdensities in a 4.5$\mu$m selected galaxy sample with robust photometric redshifts \citep{brodwin06_iss}. The selection is independent of the presence of a red sequence. There are 335 clusters and groups in the ISCS sample, identified
over 7.25 deg$^2$ within the Bo\"otes field, and 1/3 of the groups/clusters are at $z > 1$.  We have spectroscopically confirmed over 20 clusters spanning $1 < z < 1.5$.  (\citealt{stanford05,brodwin06_iss,elston06,eisenhardt08,brodwin11}, Brodwin et al. in prep, Zeimann \etal\ in prep).  

We have begun extending the work of the ISCS to more effectively target the higher redshift range by
making use of the deeper IRAC imaging obtained by the Spitzer Deep, Wide Field Survey \citep[SDWFS,][]{ashby09}.  The IRAC exposure time of SDWFS is $4 \times$ that of the IRAC data used in the ISCS, which has allowed us to better identify cluster candidates at $z > 1.5$ in the Bo\"otes field as part of the new IRAC Distant Cluster Survey (IDCS).  Here we report the first spectroscopic confirmation of one of these candidates, IDCS J1426+3508 at $z=1.75$, which we believe to be a massive cluster.  Companion papers in this volume report on the mass of this cluster (Brodwin et al. 2012) and on a gravitational arc (Gonzalez et al. 2012).  We present the optical and infrared  imaging in \textsection{\ref{Sec: img}, the spectroscopic observations and resulting redshifts in \textsection{\ref{Sec: spec}} and \textsection{\ref{Sec:redshifts}, respectively, and the X-ray observations in \textsection{\ref{Sec:xray}}.  We use Vega magnitudes and a WMAP7+BAO+$H_0$ $\Lambda$CDM cosmology \citep{komatsu11}: $\Omega_M = 0.272$, $\Omega_\Lambda = 0.728$, and $H_0 = 70.4$ km s$^{-1}$ Mpc$^{-1}$.

\section{Optical and Near-IR Imaging}
\label{Sec: img}

The cluster candidate was originally identified using the SDWFS data matched with the NDWFS optical data following the same procedures described in detail in \citet{eisenhardt08}.  A significant overdensity in the 3-dimensional space of (RA, Dec, and photometric redshift) was selected for further study.  A color image made from the NDWFS optical $+$ IRAC data is shown in the left panel of Figure~\ref{fig:img} where a tight red group of galaxies is visible.   The photometric redshift estimate of the cluster is $z \sim 1.8$.

\begin{figure*}[bthp]
\centerline{
 \mbox{\includegraphics[totalheight=0.375\textheight,width=0.5\textwidth]{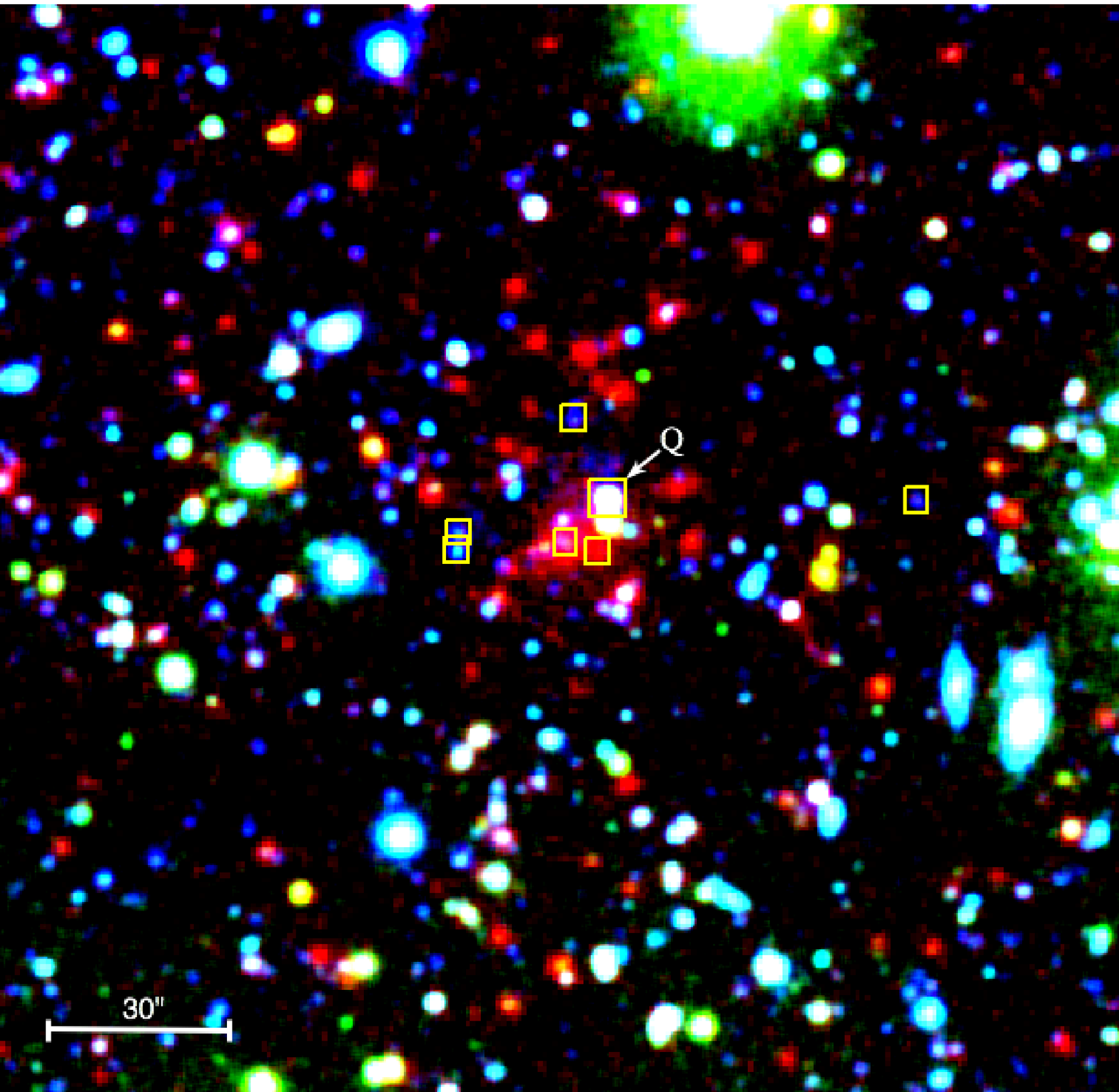}}
 \mbox{\includegraphics[totalheight=0.375\textheight,width=0.5\textwidth]{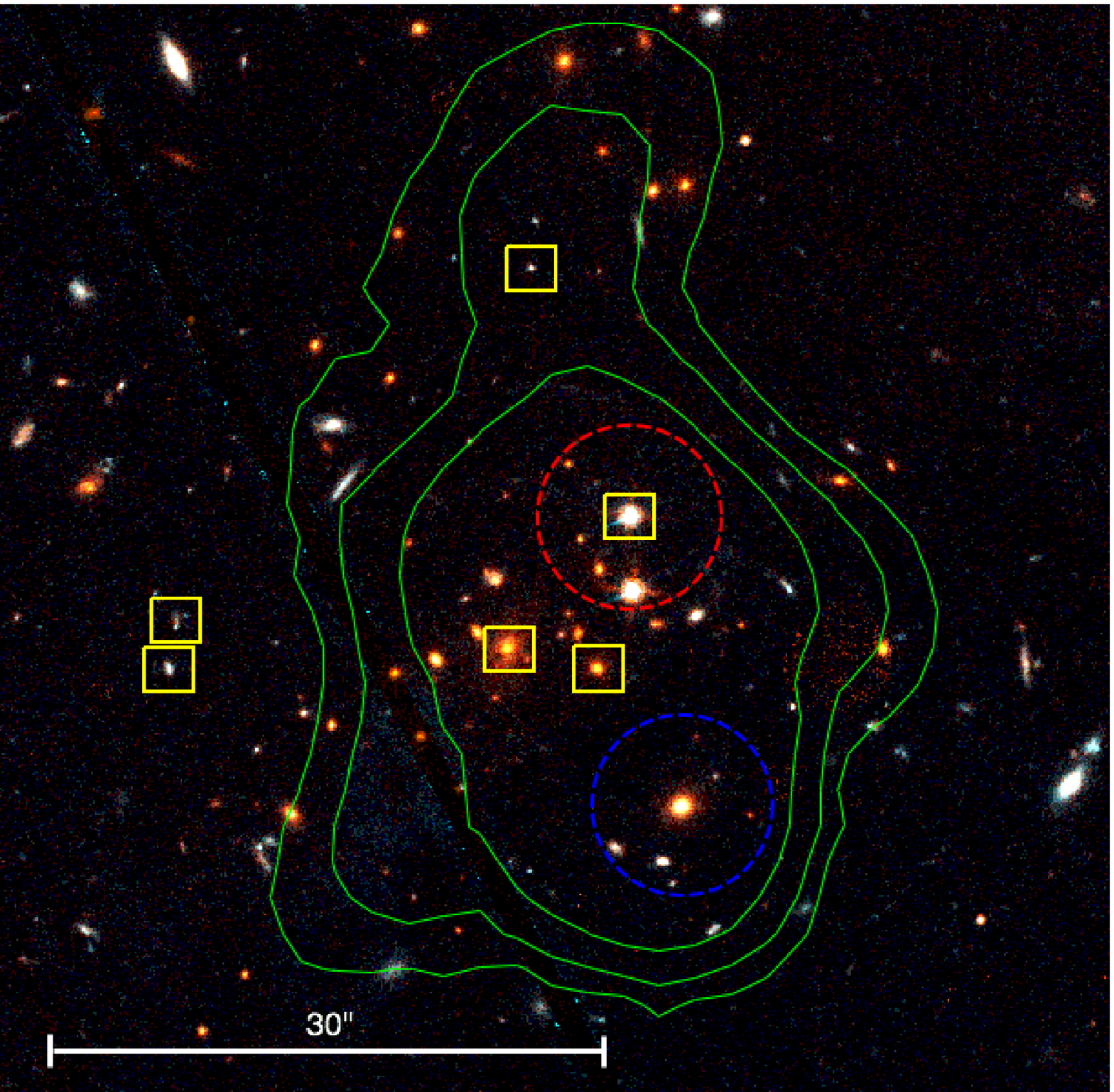}}
}
\caption{(left) Color image covering 3 
$\times$ 3 arcmin using imaging from the NDWFS $B_W$ and $I$, and 
IRAC  4.5\, $\mu$m data centered on IDCS J1426+3508.  The Q marks 
the quasar in the cluster.  (right) Pseudo-color HST image made 
from the ACS/F814W and WFC3/F160W images with the green contours 
illustrating the X-ray emission. The dashed red circle is centered 
on the quasar in the cluster, and the blue dashed circle is centered 
on a non-member radio-loud AGN. The radii of these two dashed circles 
is 5 arcsec, the same size as was used to mask these point sources 
in the X-ray analysis.  In both panels the yellow boxes are spec-z 
confirmed members, and a 30\arcsec\, (260 kpc) scale bar is given. }

\label{fig:img}
\end{figure*}

Deeper optical and NIR follow-up imaging was obtained with the 
Hubble Space Telescope using ACS and WFC3. F814W exposures were
obtained with ACS in one pointing for $8 \times 564$~s and reduced
using standard procedures.  The WFC3 data were obtained with the F160W
filter in two slightly overlapping pointings, each comprising 700~s of
integration time in dithered exposures.  The WFC3 data were reduced
using standard procedures with the MultiDrizzle software.  A pseudo-color image
constructed from the registered ACS and WFC3 imaging is shown
in the right panel of Figure~\ref{fig:img}.

Photometry was measured from the HST imaging using SExtractor 
\citep{sext} in dual image mode with sources detected in the WFC3
image.  Colors were measured in 0.8 arcsec diameter apertures, and
MAG\_AUTO is used as a``total'' magnitude.  Our photometric uncertainties are dominated by sky shot noise. We estimated the uncertainties from the distribution of sky background measurements in 5000 randomly placed 0.8\arcsec\ apertures. These sky measurements are roughly normally distributed; we estimate $\sigma_{sky}$ by fitting a Gaussian to the left-half of the distribution (which is uncontaminated by light from objects). 
We verified this procedure by confirming that it produces the correct
scaling in photometric scatter of sources detected in the sets of
dither images before making the final stack.  

Morphologies of the galaxies in the WFC3 image were determined by using Galapagos \citep{galapagos} to run GALFIT
\citep{galfit}.   Galapagos measures the sky around every galaxy and uses the basic 
isophotal parameters measured by source extractor to generate a first 
guess set of Sersi\c{c} parameters for each galaxy.  It then uses these 
values to fit a single Sersi\c{c} profile to every galaxy with GALFIT, 
simultaneously fitting close neighbors.    We use the Sersi\c{c} index $n$ to classify galaxy morphologies as being either early-type ($n>2.5$) or late-type ($n\le2.5)$.

\section{Spectroscopy}
\label{Sec: spec}

\subsection{Keck Optical Spectra}

Spectra with the Low-Resolution Imaging Spectrograph (LRIS;
\citealt{lris}) on Keck I were acquired on UT 2011 April 28 and 29
using $1.1\arcsec \times 10\arcsec$ slitlets, the G400/8500 grating on
the red side, the D680 dichroic, and the 400/3400 grism on the blue
side.  On the first night when the seeing was 0\farcs6 and conditions
were mostly clear, data were obtained in one mask in six 1200~s
exposures.  On the second night when the seeing varied in the range
$0.6 - 0.8$ arcsec with light cirrus, data were obtained in a second
mask for ten 1200~s exposures.  These data were split into slitlets
which were separately reduced following standard procedures.  The
relative spectral response was calibrated via longslit (with a width of 1.0 arcsec) observations of
Wolf 1346 and Feige 34 \citep{mg90}.

\subsection{WFC3 Grism Spectra}

The WFC3 data were obtained on 2010 November 6.  A total exposure time of 11247~s was used for the G102 observation and 2011~s for G141; a single pointing was used in both cases.  We targeted the cluster candidate with both IR prisms so as to provide continuous wavelength coverage of nearly 1~$\mu$m, allowing for conclusive identifications of spectral features in the redshift range of interest.  The FWHM of an unresolved emission line is $\sim$2 pixels, which corresponds to a spectral resolution of 49 \AA\  for G102 and 93 \AA\  for G141.  The low grism resolution blends some of the more common emission lines such as H$\alpha$+N[II] and [OIII]5007+[OIII]4959.  This typically results in a redshift uncertainty of $\sigma_z  \sim$ 0.01.   

In slitless spectroscopy, a direct image is a necessary companion to the grism image in order to zero-point the wavelength scale and to  properly extract spectra.    We chose broadband filters that closely matched the grism spectral coverage: F105W for G102 and F140W for G141.  The object positions and sizes measured from the direct images are used to establish the location, wavelength zero point, and spectral extraction widths of the objects in grism images. The data were reduced using aXe (Kummel et al. 2008).   The calibration files\footnote[1]{http://www.stsci.edu/hst/wfc3/analysis/grism\_obs/calibrations/} used were the  best available at the time of the reductions.  The steps used to extract spectra are very similar to that found in WFC3 Grism Cookbook\footnote[2]{http://www.stsci.edu/hst/wfc3/analysis/grism\_obscookbook.html} and a more detailed description will be presented in Zeimann et al. (in preparation).

\begin{figure*}[hbtp]
\includegraphics[totalheight=.55\textheight,width=1\textwidth]{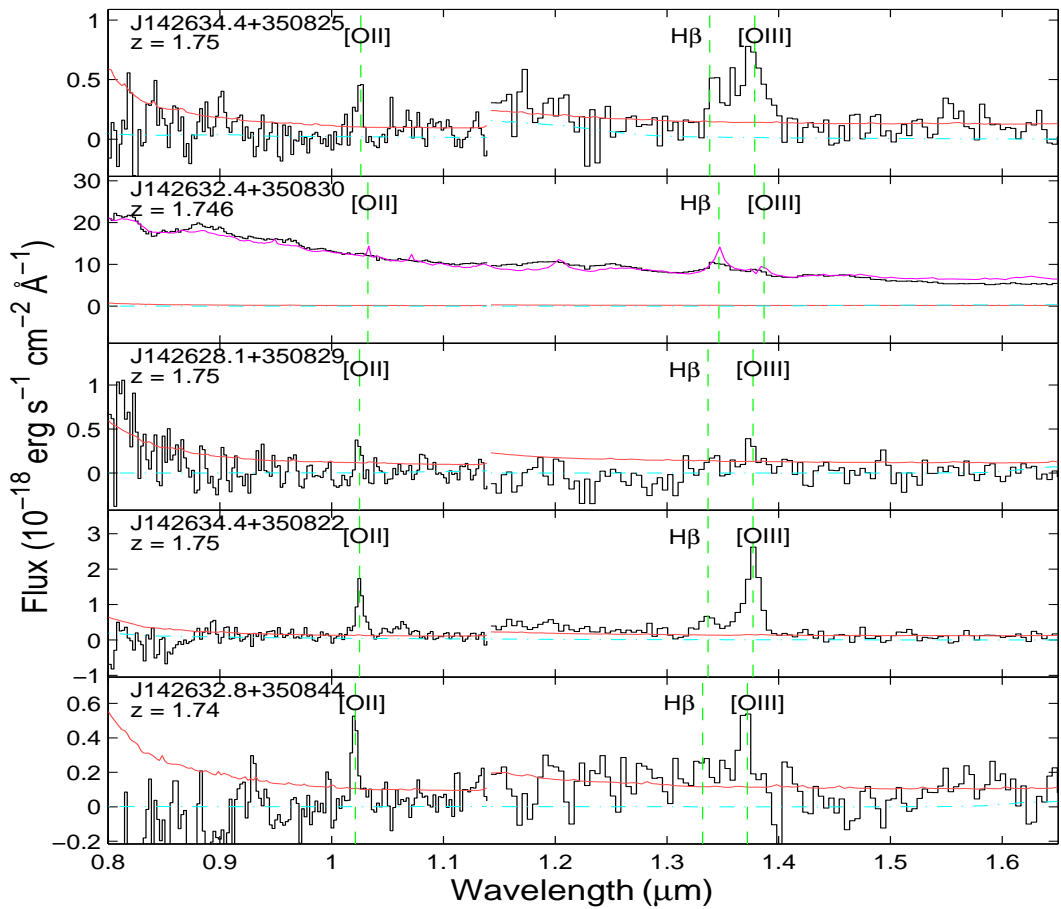}
\caption{WFC3 spectra of the five cluster members that exhibit emission lines are plotted above.  The solid black histograms are the spectra from the G102 and G141 grisms.  The dot-dashed cyan line is the estimate of contamination from overlapping spectra which is subtracted off in the final stage of reduction.  The solid red line is the 1-$\sigma$ flux error.  The vertical green lines which are labeled are the detected or expected emission from the [OII]$\lambda$3727, H$\beta$, and [OIII]$\lambda$5007 lines at the nominal cluster redshift.   The bright, power-law spectrum second from the top is a QSO, previously identified in AGES optical spectroscopy (Kochanek et al., in preparation); in this panel a QSO template (SDSS; Vanden Berk et al. 2001) is shown by the magenta line.  }
\label{fig:late}
\end{figure*}

\begin{figure*}[hbtp]
\includegraphics[totalheight=.55\textheight,width=1\textwidth]{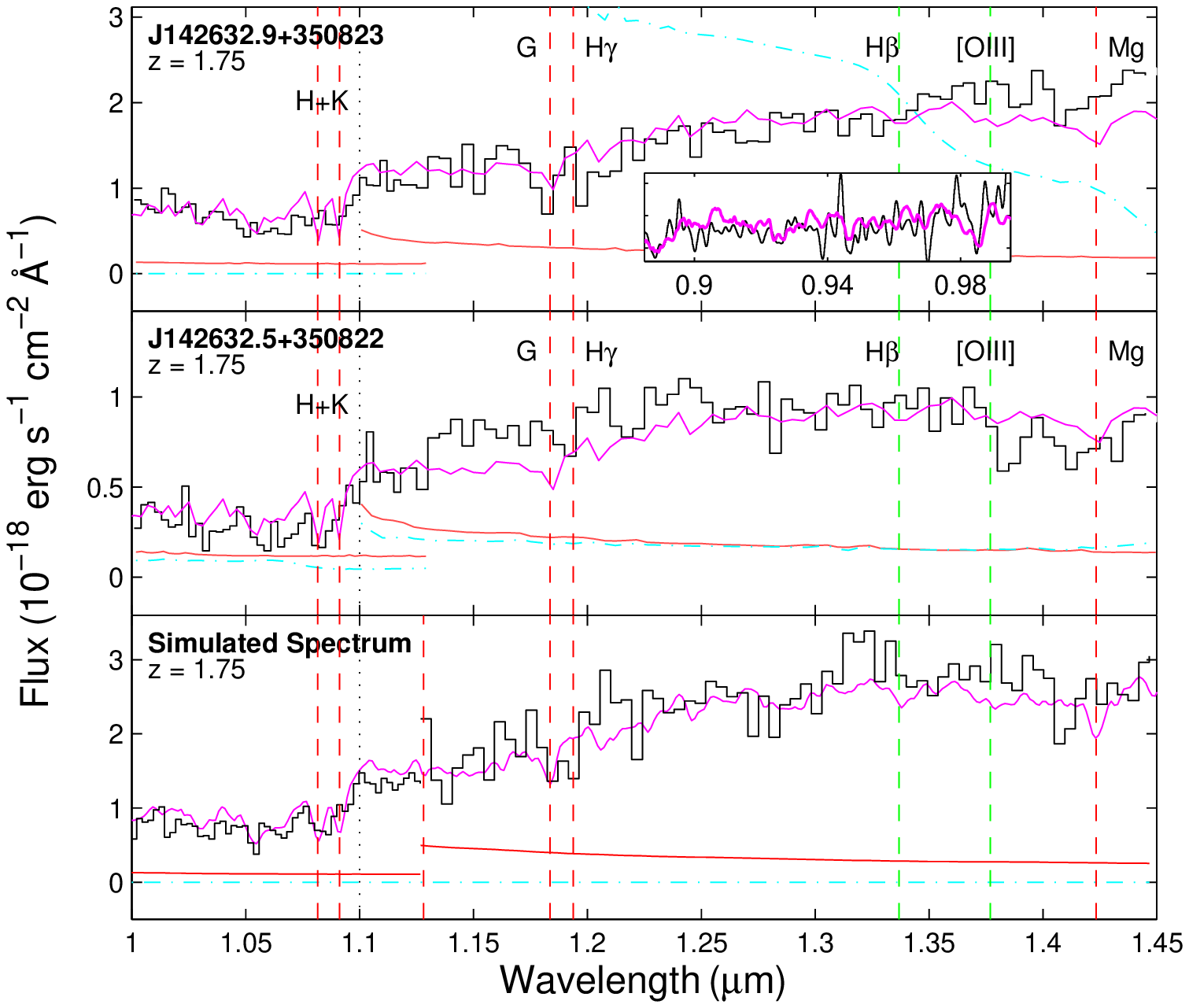}
\caption{WFC3 spectra of the two cluster members with early-type
  spectra in the top two panels; the bottom panel shows a simulated spectrum, described in the text, for reference.  The solid black histograms are the spectra from the G102 and G141 grisms.  The dot-dashed blue line is the estimation of contamination from overlapping spectra which is subtracted off in the final stage of reduction.  The solid red line is the 1-$\sigma$ flux error.  The vertical green lines are the expected locations of the [OII]$\lambda$3727, H$\beta$, and [OIII]$\lambda$5007 lines.  The vertical red lines are the expected locations for the following absorption features: Ca H+K, the G-band, H$\gamma$, and MgII$\lambda$2800.   The magenta lines represent the SDSS LRG template fitted to the observed spectra.  The inset spectrum in the top panel shows the LRIS spectrum (black solid line) in the vicinity of the D4000 break, along with the template fit (magenta solid line), solidifying the reality of this feature seen in the grism spectrum. }
\label{fig:early}
\end{figure*}

\section{Redshift Measurements}
\label{Sec:redshifts}
The reduced optical spectra were visually inspected to determine redshifts.  Despite the long integrations on a 10~m telescope using a spectrograph with new red-sensitive CCDs, sufficiently good LRIS spectra were obtained on only two phot-z selected objects which were found to have features such as D4000, B2640 and the MgII$\lambda$2800 absorption line that are characteristic of older stellar populations.  These features indicate $z = 1.75$ for both objects, which were confirmed by the WFC3 IR grism spectroscopy.

The NIR grism spectra were first visually inspected.  Emission lines were identified as being blended H$\alpha$+[NII], [OIII]$\lambda$5007$+$[OIII]$\lambda$4959, H$\beta$, or [OII]$\lambda$3727.   The spectra were also cross-correlated with spectral templates (taken from www.sdss.org/dr7/algorithms/spectemplates/index.html) to automatically determine redshifts where feasible (Zeimann et al., in prep).  Contamination from overlapping spectra was estimated using a Gaussian emission model, scaled by the measured broadband magnitudes from the direct images (F105W or F140W; see the WFC3 Grism Cookbook for more details).  This process is handled in the standard reduction package of aXe.   For the ETG spectra, we experimented with a range of SDSS galaxy templates and found that the LRG template was the best match to our grism data.

\begin{figure}[hbtp]
\centerline{
\includegraphics[totalheight=.4\textheight,width=0.66\textwidth]{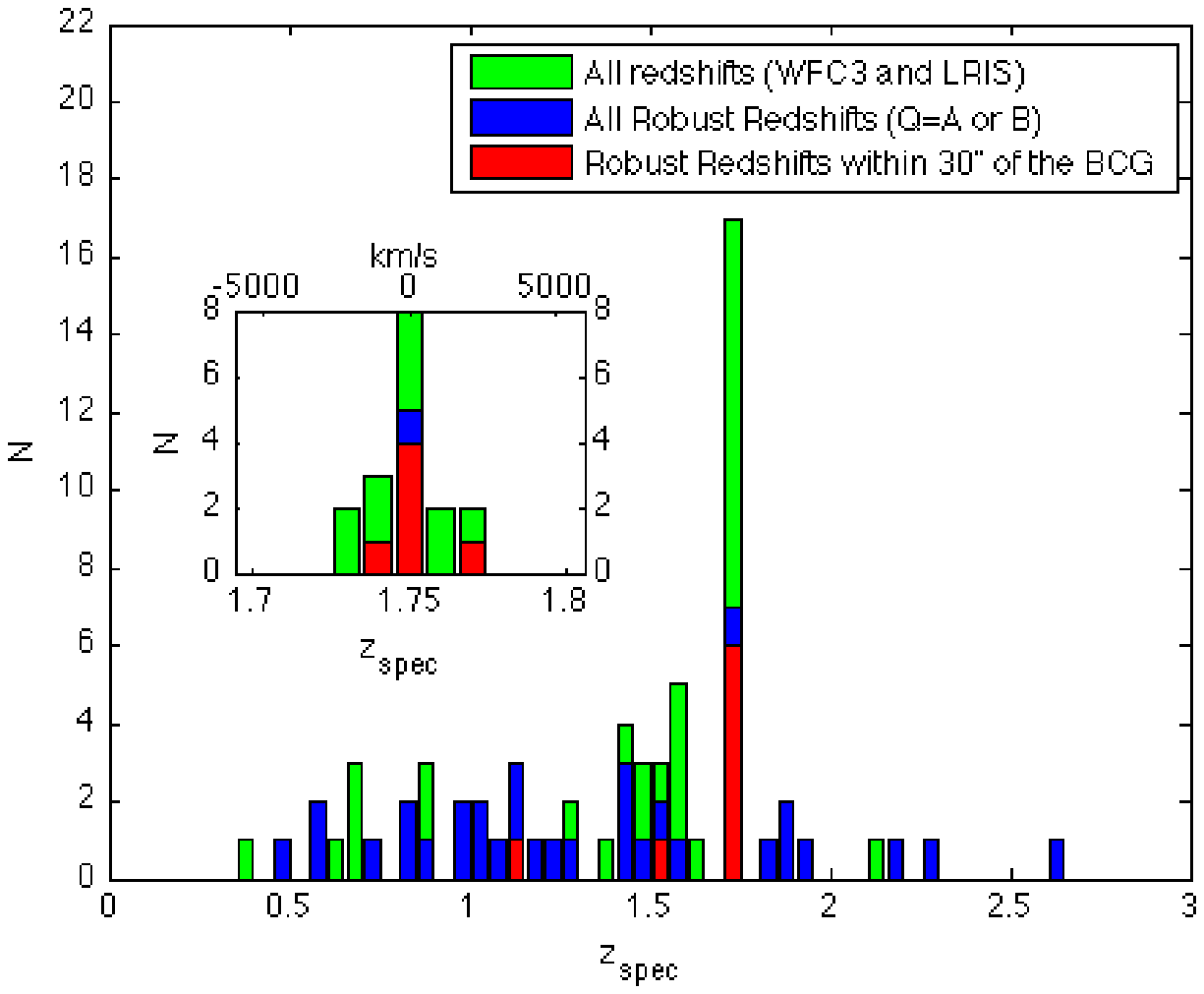}
}
\caption{Redshift histogram resulting from the LRIS and WFC3 spectroscopic observations. The green bars show all spectroscopic redshifts of all qualities. The blue histogram shows only the robust redshifts (i.e., quality A or B), and the red histogram shows the subset of these which lie within 30" of the brightest cluster galaxy. The inset shows a detail of the redshift histogram near the cluster redshift; its width is consistent with the uncertainties of the grism redshifts.}
\label{fig:zhist}
\end{figure}

The spectra of cluster members are shown in Figure~\ref{fig:late} and Figure~\ref{fig:early}.  The latter shows in the bottom panel a simulated early-type galaxy spectrum, as observed by WFC3 with our observational parameters.  aXeSim was used with an SDSS LRG template spectrum\footnote[3]{http://www.sdss.org/dr7/algorithms/spectemplates/index.html}, redshifted to $z=1.75$, to create a mock grism image.  A 1-D extraction was performed with the same reduction procedure as was used with the actual WFC3 grism observations.  The redshifts determined from both the LRIS and WFC3 spectroscopy on the identified cluster members are summarized in Table~\ref{members}, along with magnitudes and colors obtained from the HST images.  The overall redshifts obtained in the vicinity of the cluster candidate with the LRIS and WFC3 grism data are presented in Figure~\ref{fig:zhist}.

\begin{deluxetable*}{llllllccc}

\tabletypesize{\scriptsize}
\tablecaption{Spectroscopic Cluster Members}
\tablenum{1}
\tablehead{
\colhead{ID} & \colhead{R.A. (J2000)} & \colhead{Decl. (J2000)} & \colhead{$z$} & \colhead{$\Delta$z} & \colhead{Instrument} & \colhead{F160W} & \colhead{F814W-F160W}}
\startdata
 J142632.9+350823\tablenotemark{a} &  14:26:32.95  & 35:08:23.6 & 1.75 & 0.01 & WFC3/LRIS\tablenotemark{b} & 19.25 & 4.47 \\
 J142632.5+350822 &  14:26:32.55  & 35:08:22.5 & 1.75 & 0.01 & WFC3/LRIS\tablenotemark{c}  & 20.41 & 4.16 \\
 J142632.4+350830\tablenotemark{d} &  14:26:32.40  & 35:08:30.8 & 1.746\tablenotemark{e} & 0.01 & WFC3 & 18.59 & 1.08 \\
 J142632.8+350844 &  14:26:32.85  & 35:08:44.4 & 1.74 & 0.01 & WFC3  & 22.81 & 1.20 \\
 J142634.4+350825 &  14:26:34.43  & 35:08:25.1 & 1.75 & 0.01 & WFC3  & 22.63 & 1.14 \\
 J142634.4+350822 &  14:26:34.47  & 35:08:22.4 & 1.75 & 0.01 & WFC3  & 22.27 & 0.90 \\
 J142628.1+350829 &  14:26:28.15  & 35:08:29.7 & 1.75 & 0.01 & WFC3  & 23.00 & 1.43  \\
\enddata
\tablenotetext{a}{Brightest Cluster Galaxy}
\tablenotetext{b}{\lri S spectrum shows a break at 2640 \AA\ and a very red continuum consistent with $z=1.75$}
\tablenotetext{c}{\lri S spectrum shows a red continuum and a MgII$\lambda$2800 absorption feature consistent with $z=1.75$}
\tablenotetext{d}{QSO}
\tablenotetext{e}{AGES redshift; WFC3 grism redshift is 1.77}
\label{members}
\end{deluxetable*}

\section{X-ray Observations}
\label{Sec:xray}
The Bo\"otes field has been surveyed previously with ACIS-I onboard
the Chandra X-Ray Observatory \citep{murray05,kenter05,brand06}.  At the
position of IDCS J1426+3508, exposures totalling 9.5 ks are available
from the Chandra archive.  These data are split between an observation
of 4.8 ks on UT 2006 July 30 (ObsID 3621) and an observation of 4.7
ks on UT 2006 August 21 (ObsID 7381).  We processed the data following
standard procedures using the Chandra Interactive Analysis of
Observations (CIAO; V4.2) software.  We initially identified good-time
intervals for the exposures, yielding a total effective exposure
time of 8.3 ks for IDCS J1426+3508.  

The cluster is clearly detected
as an extended source in both individual exposures, as well as in
the stacked exposure.  The cluster is approximately 6.5 arcmin
off-axis in both exposures, for which the Chandra point-source 50\%
encircled energy radius is 2 arcsec at 1.5 keV.  This
complicates the X-ray analysis, as an optically bright quasar
confirmed to be in the cluster (see Section 5) is only 9 arcsec from the brightest cluster
galaxy (BCG).  In addition, we identify an X-ray point source
associated with a radio source that is 12 arcsec to the SW, at
14:26:32.2, +35:08:14.9.  This source, confirmed as an emission line galaxy at
$z = 1.535$ in our WFC3 grism data, has an integrated 21~cm flux
density of 95.3~mJy from the FIRST survey \citep{becker95}.  Given the signal to
noise ratio and the large off-axis angle of the available X-ray observations,
it is challenging to disentangle the extended cluster emission from
the point source contributions.   However, as seen in the right panel of Figure~\ref{fig:img}, IDCS
J1426+3508 is clearly associated with diffuse
X-ray emission that extends beyond the point-source contributions
from the two AGN.  We see no evidence in the WFC3 grism spectroscopy, which covers the central 2 arcmin of the cluster, for other AGN which could contribute to the X-ray emission.  We have also used the SDWFS IRAC photometry to construct a two-color diagram to search for obscured AGN (following Stern et al. 2005) which might contribute to the measured X-ray flux.  In addition to the QSO and the radio source already described, there is one more object within the X-ray measurement aperture which has IRAC colors typical of obscured AGN.  

To extract the X-ray counts due to the cluster, we masked the three 
AGN (the QSO, the radio source, and the IRAC AGN) using a conservative
5 arcsec radius aperture, corresponding to the 90\% encircled energy
radius at the observed off-axis-angle of the cluster.  We expect that the unmasked flux from the two AGN to contribute only one photon to the measured flux. We then
extracted cluster source counts in the 0.5 - 7 keV range within a 1
arcmin radius aperture centered on the cluster BCG.  This aperture
approximately corresponds to a radius of 500 kpc at the cluster redshift.
Response matrices and effective areas were then determined for each
detected source.  Within the measurement aperture, there are $54 \pm 10$ background-corrected counts in the $0.5-7$ keV range, after masking out the two central AGN.  We used XSPEC (V12.6.0) to fit the background-subtracted
X-ray spectrum with the MEKAL hot, diffuse gas model \citep{mewe85} using the Wisconsin photo-electric absorption cross-section
\citep{mm83}.  The temperature was fixed at 5 keV and the abundance at 0.3 M$_\odot$, with 
a Galactic absorption of 1.3 $\times 10^{20}\, {\rm cm}^{-2}$ at 
the target position.  We determined a Galactic absorption-corrected flux of $(3.1 \pm 0.7)
\times 10^{-14}$ ergs cm$^{-2}$ s$^{-1}$ in the $0.5 - 2.0$ keV range, which
translates to an X-ray luminosity of $(5.5 \pm 1.2) \times 10^{44}$
ergs s$^{-1}$ at $z=1.75$.   The X-ray flux changes by only 7\% if the X-ray temperature is varied from 4 to 6 keV. 

Using the M$_{500} - $L$_X$ relation of \citet{vikh09}, we estimate from 
the luminosity that $M_{500, L_X} = (3.5 \pm 1.0) \times10^{14}$ \msun. We 
caution that the use of this scaling relation requires a significant 
extrapolation in redshift.  This and other systematic 
uncertainties (such as removal of X-ray point sources) are expected to 
dominate over statistical errors.  To estimate a total cluster mass, 
we next assume an NFW density profile \citep{navarro97} and the 
mass-concentration relation of \citep{duffy08}. The resulting 
$M_{200} = (5.6 \pm 1.6) \times10^{14}$ \msun, where the uncertainty is determined only from the statistical measurement error.

\section{Galaxy Populations}

Figure~\ref{fig:cmd} presents the photometric and morphological information for all the objects in the 7.7 arcmin$^2$ area centered on the cluster where the ACS and WFC3 imaging overlap.  
While it is possible to see a red sequence of early-type galaxies in Figure~\ref{fig:cmd}, the spread and location is different from that of the red sequence in massive clusters at $1 < z < 1.5$.   
To isolate a sample of probable cluster members in the color-magnitude diagram we carry out the following multi-step procedure.  
First, we determined the relative offset
$\Delta$ between the object colors and the expected color-magnitude relation at this
redshift (based on a model for Coma with $z_f=3$; see \citet{eisenhardt07} for
details). The range $-0.5 < \Delta < 1.0$ (corresponding to colors
$2.3<$ F814W $-$ F160W $< 3.8$) is the color cut used to select member galaxies on the red
sequence. 
We initially define the cluster red sequence as galaxies brighter than $H^*(z) + 1.5$ that are chosen by the above color cut in $\Delta$, where $H^*(z)$ is passively evolved from the luminosity of Coma early-type galaxies.  
Most of the objects in this color-selected red sequence do not
have spectroscopic or photometric redshift information,
and so the initial red sequence sample may suffer from interloper contamination. For the purpose of studying the
cluster red sequence, we choose to restrict consideration to morphologically selected early-type galaxies, keeping objects with $n > 2.5$.  We discard objects with colors
more than two absolute deviations from the central red sequence color, where the deviation is the median of the absolute value of the $\Delta$ of the potential red sequence objects.  Objects removed in
the latter step are shown with a gray dot in Figure~\ref{fig:analysis}.   The remaining objects in the red sequence sample are represented by the red points in Figure~\ref{fig:analysis}.

\begin{figure*}[bthp]
\includegraphics[totalheight=0.35\textheight,width=1.0\textwidth]{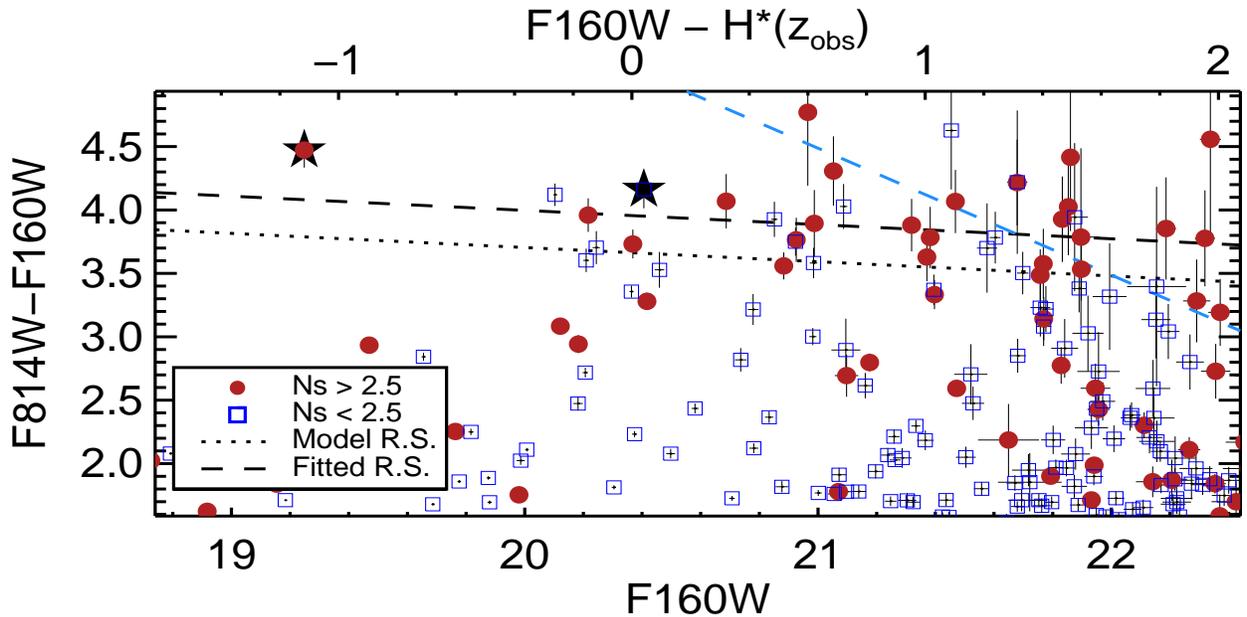}
\caption{Color-magnitude diagram of IDCS~J1426+3508, made from ACS and
  WFC3 imaging.  The red and blue colored points denote morphological classifications, based on the Sersi\c{c} index.  The spectroscopically confirmed members are marked by the larger stars; only two of the members are red enough to appear in this plot.   The black dotted line represents the expected color of a
  passively evolving red sequence of galaxies formed at $z_f = 3$, with the slope based on observed Coma colors
  \citep{eisenhardt07}.  The diagonal black dashed line is the fit to the red sequence galaxies.  The diagonal dashed blue line represents
  the 5 $\sigma$ limit on the colors.  }

\label{fig:cmd}
\end{figure*}

To measure the color and scatter of this red sequence
sample down to F160W $= 21.86$, we use the biweight estimates of location and
scale  \citep{beers90}. We calculate the intrinsic scatter $\sigma_{int}$ in the red sequence sample by subtracting in quadrature the
median color error from the biweight scale estimate.  The red horizontal
lines in Figure~\ref{fig:analysis} are offset a distance $\sigma_{int}$ above and below the
median color of the red sequence.  We
estimate uncertainties by performing these calculations
on 1000 bootstrap resamplings, from which we measure
the scale of the resulting color and $\sigma_{int}$ distributions using their median absolute deviation. The intrinsic scatter in the observed colors of the red sequence galaxies is
$0.16 \pm 0.06$ mag; transformed to the rest frame $U - V$ the
intrinsic scatter is $0.10 \pm 0.04$. The median color of the morphologically-selected 
red sequence sample is F814W $-$ F160W $= 3.96 \pm 0.07$, which
is $0.29 \pm 0.05$ redder than the expected color, derived from a 
simple passive evolution model with
a $z_f = 3$  \citep[2007 version]{bc03}.  A model with an earlier formation epoch, $z_f = 6$, predicts a color which is closer to the observed F814W $-$ F160W colors of the morphologically-selected red sequence.

\begin{figure*}[bthp]
\includegraphics[totalheight=0.36\textheight,width=1.0\textwidth]{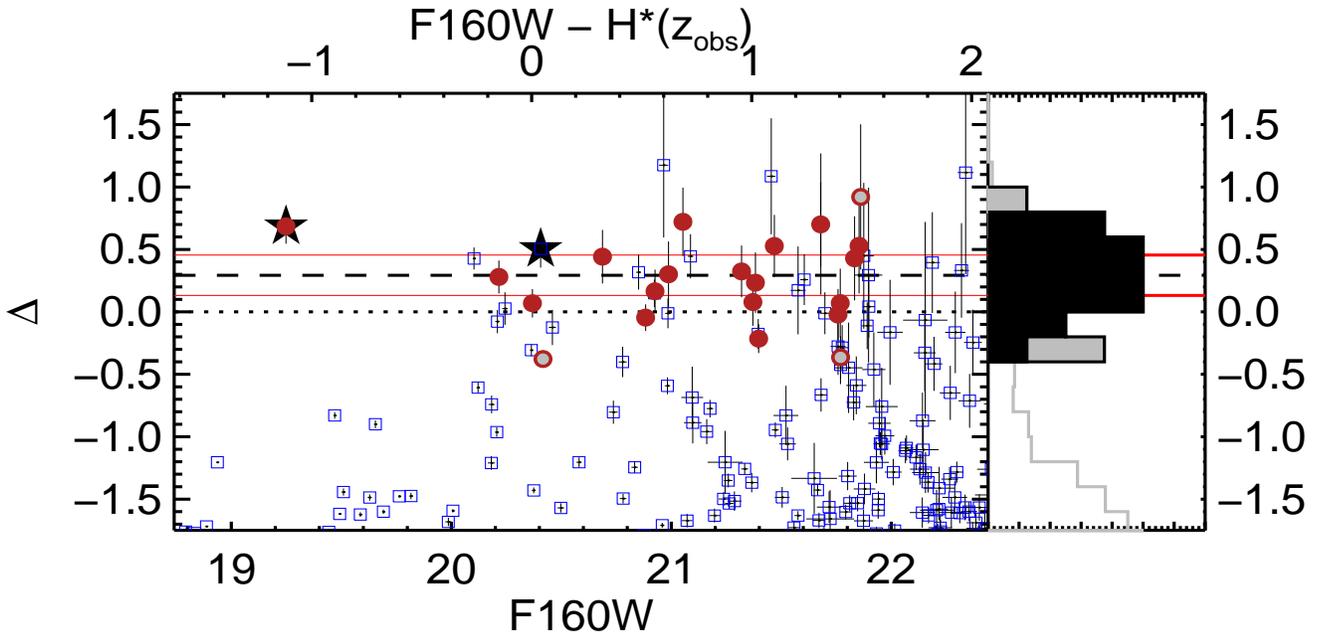}
\caption{The left panel of the plot is similar to the color-magnitude diagram shown in Figure~\ref{fig:cmd}.  The F814W $-$ F160W colors have been zero pointed to the color predicted by a
  passively evolving red sequence of galaxies formed at $z_f = 3$, with the slope based on observed Coma colors
  \citep{eisenhardt07}.  The red filled points are the objects selected by the magnitude and color cuts described in the text which have a Sersi\c{c} index $n > 2.5$.  The grey-filled circles were removed from the red sequence selection.  The solid red horizontal lines show the one $\sigma_{int}$ intrinsic scatter of the red sequence galaxies, and the black dashed line is the fit to the colors of these galaxies. The two spectroscopically confirmed members that lie within the limited color range of this CMD are marked by the larger stars.  The right side of the plot is a histogram stacked in the colors.  }

\label{fig:analysis}
\end{figure*}

The cluster includes a fairly large brightest cluster galaxy.  Profile fitting using GALFIT determined a Sersi\c{c} index of $5.4 \pm 0.1$, $r_e = 18.0  \pm 2.5$ kpc, and a total F160W magnitude $= 18.5 \pm 0.05$ which is approximately 2.0 magnitudes brighter than $M^*$ for a passively-evolving galaxy formed at $z_f = 3$ \citep{bc03}.  From the F160W total magnitude, we determine a rest frame $M_V = -24.7$ assuming the $z_f = 3$ passive evolution model to transform to rest V-band.  The size and luminosity of this BCG is similar to a BCG at $z \sim 0$, which is remarkable since it is at $z = 1.75$.


\section{Discussion}
\label{Sec: discussion}

IDCS~J1426+3508 is a newly discovered galaxy cluster at $z=1.75$, which places it among the few such systems currently known at $1.5 < z < 2$, when massive clusters may be first forming.  Using a combination of optical multi-object Keck spectroscopy and infrared \hst/WFC3 grism spectroscopy, we have confirmed 7 cluster members in IDCS~J1426+3508 within a radius of 2\,Mpc, all but one of which are within a radius of 250 kpc.  The extended X-ray emission described in Section~4 indicates that this cluster is gravitationally bound and already has a mass greater than $10^{14}$ M$_\odot$. 

The properties of the galaxies in the cluster indicate that the cluster is far from settled in terms of star formation.  Most of the spectroscopic member galaxies are very blue (c.f. Table~\ref{members}) and show emission lines in their spectra indicative of on-going star formation or AGN activity.  The red sequence itself has a larger amount of scatter, $0.16$ mag, compared to $0.08$ in a similar observed color in XMMU~J2235$-$2557, a slightly more massive cluster at somewhat lower redshift, $z = 1.39$ \citep{strazz10}.  The intrinsic scatter in IDCS~J1426+3508 is reasonable for a population of galaxies whose stars formed at $z_f \sim 3.5 - 6$.   The median color of these galaxies matches the prediction for simple passive evolution if the stars in the red sequence galaxies were formed at $5 < z_f < 6$  \citep[2007 version]{bc03}.  This is in contrast to the cluster at $z=1.62$ \citep{pap10, tan10} in which the red sequence has a younger $z_f$.  The red sequences may have different median colors because of the difference in the cluster masses, in that IDCS J1426+3508 is more massive and so may have formed earlier.   Reddening by dust could cause the redder colors in IDCS~J1426+3508, though 
there is no reason to suspect unusual amounts of dust in this cluster.  Indeed, most of the spectroscopically confirmed members are fairly blue.

\section{Conclusions}
\label{Sec: conclusions}

We have presented optical and NIR spectroscopy of galaxies in IDCS~J1426+3508, an IRAC-selected galaxy cluster candidate, which
indicate this is a bona fide cluster of galaxies at $z = 1.75$.  
The available Chandra data show a faint but clearly extended X-ray source at this location.   
Along with the centrally concentrated and regular distribution of red galaxies, the X-ray detection argues that IDCS~J1426+3508 is gravitationally bound and in a relatively relaxed state.   

The discovery of a cluster at $z = 1.75$ with a soft X-ray flux
greater than the $\sim 10^{-14}$ ergs cm$^{-2}$ s$^{-1}$ limit of the upcoming
eROSITA mission \citep{erosita}, scheduled for launch
in 2014, promises exciting results for that mission.  However, we
note that at the 28 arcsec average resolution expected for the
eROSITA slew survey, and even at the 15 arcsec on-axis resolution
of eROSITA, it would have been very difficult to distinguish the AGN from the
cluster emission.  Without the AGN masked out in our Chandra data,
the derived X-ray flux for IDCS~J1426+3508 would be $\sim 50\%$
higher.  As shown by e.g. \citet{galametz09},
the fraction of galaxy clusters that host luminous AGN increases
rapidly with redshift.  Such AGN will make it challenging to draw
firm cosmological evolutionary results from the eROSITA cluster
sample without higher resolution X-ray follow-up.

With the higher resolution Chandra data currently available from the archive, the point-source corrected X-ray
luminosity $L_{0.5-2.0 keV} = (5.5 \pm 1.2) \times 10^{44}$ ergs s$^{-1}$, which implies  M$_{200,L_x} \sim 5 \times 10^{14}$~M$_\odot$. This is a surprisingly large cluster mass for this redshift and survey area, as will be addressed in forthcoming papers.

\acknowledgments This work is based in part on observations obtained
with the Chandra X-ray Observatory (CXO), under contract
SV4-74018, A31 with the Smithsonian Astrophysical Observatory which
operates the CXO for NASA.  Support for this research was provided by
NASA grant G09-0150A.  AHG acknowledges support from the National Science Foundation through 
grant AST-0708490.  This work is based in part on observations made
with the Spitzer Space Telescope, which is operated by the Jet
Propulsion Laboratory, California Institute of Technology under a
contract with NASA. Support for this work was provided by NASA through
an award issued by JPL/Caltech.  Support for HST programs 11663 and 12203
were provided by NASA through a grant from the Space Telescope Science
Institute, which is operated by the Association of Universities for
Research in Astronomy, Inc., under NASA contract NAS 5-26555.  Some of
the data presented herein were obtained at the W.~M.~Keck Observatory,
which is operated as a scientific partnership among the California
Institute of Technology, the University of California and the National
Aeronautics and Space Administration.  The Observatory was made
possible by the generous financial support of the W.~M.~Keck
Foundation.  This work makes use of image data from the NOAO Deep
Wide--Field Survey (NDWFS) as
distributed by the NOAO Science Archive. NOAO is operated by the
Association of Universities for Research in Astronomy (AURA), Inc.,
under a cooperative agreement with the National Science Foundation.

We thank Matt Ashby for creating the IRAC catalogs for SDWFS, Buell Jannuzi for his work on the NDWFS, Michael Brown for combining the NDWFS with SDWFS catalogs, Steve Murray and his XBo\"otes team for obtaining the Chandra data in the Bo\"otes field, and Alexey Vihklinin for advice on the analysis of the Chandra data.  This paper would not have been possible without the efforts of the support staffs of the Keck Observatory, 
Spitzer Space Telescope, Hubble Space Telescope, and Chandra X-ray Observatory.  
We also thank the anonymous referee for comments which helped to improve the presentation of our results. 
Support for MB was provided by the W.~M.~Keck Foundation.  The work by SAS at LLNL was
performed under the auspices of the U.~S.~Department of Energy under
Contract No. W-7405-ENG-48.


\begin{thebibliography}{}

\bibitem[\protect\astroncite{{Ashby} et~al.}{2009}]{ashby09}
{Ashby}, M.~L.~N., et al.  2009,\newblock {\em \apj,} {\bf 701}, 428

\bibitem[\protect\astroncite{{Becker} et~al.}{1995}]{becker95}
{Becker}, R., et al.  1995,\newblock {\em \apj,} {\bf 450}, 559

\bibitem[\protect\astroncite{{Beers} et~al.}{1990}]{beers90}
{Beers}, T., et al.  1990,\newblock {\em \aj,} {\bf 100}, 32

\bibitem[\protect\astroncite{{Bertin \& Arnouts}}{1996}]{sext}
{Bertin}, E., and {Arnouts}, S. 1996,\newblock {\em A\&A,} {\bf 117}, 393

\bibitem[\protect\astroncite{{Bower} et~al.}{1994}]{bower94}
{Bower}, R.~G., {Bohringer}, H., {Briel}, U.~G., {Ellis}, R.~S., {Castander},
  F.~J., and {Couch}, W.~J. 1994,\newblock {\em \mnras,} {\bf 268}, 345

\bibitem[\protect\astroncite{{Brand} et~al.}{2006}]{brand06}
{Brand}, K., et al.  2006,\newblock {\em \apj,} {\bf 641}, 140

\bibitem[\protect\astroncite{{Brodwin} et~al.}{2006}]{brodwin06_iss}
{Brodwin}, M., et al.  2006,\newblock {\em \apj,} {\bf 651}, 791

\bibitem[\protect\astroncite{{Brodwin} et~al.}{2007}]{brodwin07}
{Brodwin}, M., {Gonzalez}, A.~H., {Moustakas}, L.~A., {Eisenhardt}, P.~R.,
  {Stanford}, S.~A., {Stern}, D., and {Brown}, M.~J.~I. 2007,\newblock {\em \apjl,} {\bf 671}, L93

\bibitem[\protect\astroncite{{Brodwin} et~al.}{2011}]{brodwin11}
{Brodwin}, M., et al.  2011,\newblock {\em \apj,} {\bf 732}, 33

\bibitem[\protect\astroncite{{Bruzual \& Charlot}}{2003}]{bc03}
{Bruzual}, G., and {Charlot}, S. 2003,\newblock {\em \mnras,} {\bf 344}, 1000

\bibitem[\protect\astroncite{{Capak} et~al.}{2011}]{capak11}
{Capak}, P., et al.  2011,\newblock {\em Nature,} {\bf 470}, 233

\bibitem[\protect\astroncite{{Duffy} et~al.}{2008}]{duffy08}
{Duffy}, A.R., et al.  2008,\newblock {\em MNRAS,} {\bf 390}, L64

\bibitem[\protect\astroncite{{Eisenhardt} et~al.}{2004}]{iss}
{Eisenhardt}, P.~R.~M., et al.  2004,\newblock {\em \apjs,} {\bf 154}, 48

\bibitem[\protect\astroncite{{Eisenhardt} et~al.}{2007}]{eisenhardt07}
{Eisenhardt}, P.~R.~M., et al.  2007,\newblock {\em \apjs,} {\bf 169}, 225

\bibitem[\protect\astroncite{{Eisenhardt} et~al.}{2008}]{eisenhardt08}
{Eisenhardt}, P.~R.~M., et al.  2008,\newblock {\em \apj,} {\bf 684}, 905

\bibitem[\protect\astroncite{{Elston} et~al.}{2006}]{elston06}
{Elston}, R.~J., et al.  2006,\newblock {\em \apj,} {\bf 639}, 816

\bibitem[\protect\astroncite{{Fassbender} et~al.}{2011}]{fass11}
{Fassbender}, R., et al.  2011,\newblock {\em A\&A,} {\bf 527}, L10

\bibitem[Fruscione et al.(2006)]{ciao} Fruscione, A., et 
al.\ 2006, \procspie, 6270

\bibitem[\protect\astroncite{{Galametz} et~al.}{2009}]{galametz09}
{Galametz}, A., et al.  2009,\newblock {\em \apj,} {\bf 694}, 1309

\bibitem[\protect\astroncite{{Gobat} et~al.}{2011}]{gobat11}
{Gobat}, G., et al.  2011,\newblock {\em A\&A,} {\bf 526}, 133

\bibitem[\protect\astroncite{{Haiman} et~al.}{2001}]{haiman01}
{Haiman}, Z., et al.  2001,\newblock {\em \apj,} {\bf 553}, 545

\bibitem[\protect\astroncite{{Haussler} et~al.}{2011}]{galapagos}
{Haussler}, B., et al. 2011,\newblock {\em ASPC,}{\bf 442}, 155

\bibitem[\protect\astroncite{{Henry} et~al.}{2010}]{henry10}
{Henry}, J.~P., et al.  2010,\newblock {\em \apjl,} {\bf 725}, 615

\bibitem[\protect\astroncite{{Holder} et~al.}{2001}]{holder01}
{Holder}, G., et al.  2001,\newblock {\em \apj,} {\bf 560}, L111

\bibitem[\protect\astroncite{{Jannuzi} \& {Dey}}{1999}]{ndwfs99}
{Jannuzi}, B.~T. and {Dey}, A. 1999,\newblock in {\em ASP Conf. Ser. 191 --- Photometric Redshifts and the
  Detection of High Redshift Galaxies}, p. 111
  
\bibitem[\protect\astroncite{{Kenter} et~al.}{2005}]{kenter05}
{Kenter}, M., et al.  2005,\newblock {\em \apjs,} {\bf 161}, 9

\bibitem[\protect\astroncite{{Komatsu} et~al.}{2011}]{komatsu11}
{Komatsu}, E., et al.  2011,\newblock {\em \apjs,} {\bf 192}, 18

\bibitem[\protect\astroncite{{Kubo} et~al.}{2007}]{kubo07}
{Kubo}, J.~M., {Stebbins}, A., {Annis}, J., {Dell'Antonio}, I.~P., {Lin}, H.,
  {Khiabanian}, H., and {Frieman}, J.~A. 2007,\newblock {\em \apj,} {\bf 671}, 1466

\bibitem[\protect\astroncite{{K{\"u}mmel} et~al.}{2009}]{aXe}
{K{\"u}mmel}, M., {Walsh}, J.~R., {Pirzkal}, N., {Kuntschner}, H., and
  {Pasquali}, A. 2009,\newblock {\em \pasp,} {\bf 121}, 59

\bibitem[\protect\astroncite{{Massey} \& {Gronwall}}{1990}]{mg90}
{Massey}, P., \& {Gronwall}, C., 1990,\newblock {\em \apj,} {\bf 358}, 344

\bibitem[\protect\astroncite{{Matarrese} et~al.}{2000}]{mat00}
{Matarrese}, S., et al.  2000,\newblock {\em \apj,} {\bf 541}, 10

\bibitem[\protect\astroncite{{Mewe} et~al.}{1985}]{mewe85}
{Mewe}, R., {Gronenschild}, E.~H.~B.~M., and {van den Oord}, G.~H.~J. 1985,\newblock {\em \aaps,} {\bf 62}, 197

\bibitem[\protect\astroncite{{Morrison} \& {McCammon}}{1983}]{mm83}
{Morrison}, R. and {McCammon}, D. 1983,\newblock {\em \apj,} {\bf 270}, 119

\bibitem[\protect\astroncite{{Mortonson} et~al.}{2011}]{mort11}
{Mortonson}, M.J., et al.  2011,\newblock {\em PhysRevD,} {\bf 83}, 23015

\bibitem[\protect\astroncite{{Murray} et~al.}{2005}]{murray05}
{Murray}, S.~S., et al.  2005,\newblock {\em \apjs,} {\bf 161}, 1

\bibitem[\protect\astroncite{{Navarro} et~al.}{1997}]{navarro97}
{Navarro}, J., et al.  1997,\newblock {\em \apj,} {\bf 490}, 493

\bibitem[\protect\astroncite{{Oke} et~al.}{1995}]{lris}
{Oke}, J.~B., et al.  1995,\newblock {\em \pasp,} {\bf 107}, 375

\bibitem[\protect\astroncite{{Papovich} et~al.}{2010}]{pap10}
{Papovich}, C., et al.  2010,\newblock {\em \apj,} {\bf 716}, 1503

\bibitem[\protect\astroncite{{Peng} et~al.}{2010}]{galfit}{Peng},
  C.Y., et al. 2010,\newblock {\em \aj,} {\bf 139}, 2097

\bibitem[\protect\astroncite{{Pentericci} et~al.}{2000}]{pent00}
{Pentericci}, L., et al.  2000,\newblock {\em A\&A,} {\bf 361}, L25

\bibitem[\protect\astroncite{{Predehl} et~al.}{2010}]{erosita}
{Predehl}, P., et al.  2010,\newblock {\em arXiv:1004.5219v2}

\bibitem[\protect\astroncite{{Santos} et~al.}{2011}]{santos11}
{Santos}, J.~S., et al.  2011,\newblock {\em A\&A,} {\bf 531}, L15

\bibitem[\protect\astroncite{{Stanford} et~al.}{2005}]{stanford05}
{Stanford}, S.~A., et al.  2005,\newblock {\em \apjl,} {\bf 634}, L129

\bibitem[\protect\astroncite{{Stanford} et~al.}{2006}]{stanford06}
{Stanford}, S.~A., et al.  2006,\newblock {\em \apjl,} {\bf 646}, L13

\bibitem[\protect\astroncite{{Strazzullo} et~al.}{2010}]{strazz10}
{Strazzullo}, V., et al.  2010,\newblock {\em A\&A,} {\bf 524}, 17

\bibitem[\protect\astroncite{{Tanaka} et~al.}{2010}]{tan10}
{Tanaka}, M., {Finoguenov}, A., and {Ueda}, Y. 2010,\newblock {\em \apjl,} {\bf 716}, L152

\bibitem[\protect\astroncite{{Vanden Berk} et~al.}{2001}]{vanden01}
{Vanden Berk}, B., et al.  2001,\newblock {\em \aj,} {\bf 122}, 549

\bibitem[\protect\astroncite{{Venemans} et~al.}{2007}]{ven07}
{Venemans}, B., et al.  2007,\newblock {\em \apj,} {\bf 716}, 1503

\bibitem[\protect\astroncite{{Vikhlinin} et~al.}{2009}]{vikh09}
{Vihklinin}, A., et al.  2009,\newblock {\em \apj,} {\bf 692}, 1033

\end{thebibliography}

\clearpage

\end{document}